\documentclass[reprint,superscriptaddress,amssymb,amsmath,aps,prx,longbibliography, nofo otinbib]{revtex4-2}

\usepackage[margin= 1 in]{geometry}
\usepackage{graphicx}
\usepackage{amssymb}
\usepackage{epstopdf}
\usepackage{upgreek}
\usepackage{dcolumn}
\usepackage{bm}
\usepackage{textcomp, gensymb}  
\usepackage{braket}
\usepackage{amsmath}
\usepackage{mathtools}
\usepackage{float}
\usepackage{tabularx}
\usepackage{mathrsfs}
\usepackage{lineno}
\usepackage{tikz}
\usepackage{graphicx}
\usepackage{sidecap} % side captions
\usepackage{array}
\usepackage{amsthm}
\usepackage{enumitem}
\usepackage{xcolor} % used for flags
\usepackage{subcaption}

\usepackage[newcommands]{ragged2e}
\usepackage{caption}
\usepackage{relsize}

\expandafter\ifx\csname package@font\endcsname\relax\else
\expandafter\expandafter
\expandafter\usepackage
\expandafter\expandafter
\expandafter{\csname package@font\endcsname}%
\fi

\newcommand{\be}{\begin{eqnarray}}
\newcommand{\ee}{\end{eqnarray}}
\newcommand{\bfig}{\begin{figure}}
\newcommand{\efig}{\end{figure}}

\DeclareFontFamily{U}{mathb}{}
\DeclareFontShape{U}{mathb}{m}{n}{
  <-5.5> mathb5
  <5.5-6.5> mathb6
  <6.5-7.5> mathb7
  <7.5-8.5> mathb8
  <8.5-9.5> mathb9
  <9.5-11.5> mathb10
  <11.5-> mathbb12
}{}

\usepackage[colorlinks=true,allcolors=blue]{hyperref}%

\usepackage{blindtext}
\usepackage{subfiles} % Best loaded last in the preamble

\begin{document}

\title{Can slow recombination in ordered superconductors explain the excess quasiparticle population?}

\author{Eva Gurra}
\email{eva.gurra@colorado.edu}
\affiliation{National Institute of Standards and Technology, Boulder, CO 80305}
\affiliation{Department of Physics, University  of Colorado, Boulder, CO 80309}

\author{Douglas A. Bennett}
\affiliation{National Institute of Standards and Technology, Boulder, CO 80305}

\author{Shannon M. Duff}
\affiliation{National Institute of Standards and Technology, Boulder, CO 80305}

\author{Michael R. Vissers}
\affiliation{National Institute of Standards and Technology, Boulder, CO 80305}

\author{Joel N. Ullom}
\email{joel.ullom@nist.gov}
\affiliation{National Institute of Standards and Technology, Boulder, CO 80305}
\affiliation{Department of Physics, University  of Colorado, Boulder, CO 80309}

\date{\today}

\begin{abstract}
    \noindent An excess density of quasiparticles is widely observed in superconducting films. This excess causes performance degradation in a variety of superconducting devices, including decoherence in qubits. In this Letter, we evaluate the hypothesis of \cite{Bespalov_2016} that the quasiparticle excess is caused by anomalously slow recombination at low quasiparticle densities due to localization in sub-gap states. We probe the density of states in aluminum and niobium films using current-voltage measurements of tunnel junctions and extract upper bounds on the energy scales of the sub-gap states and gap smearing. With these parameters, we evaluate the recombination times predicted by \cite{Bespalov_2016} and find that slow recombination is not predicted to occur at observed quasiparticle densities in aluminum- and niobium-based superconducting devices. 
    These results suggest that the quasiparticle excess in ordered superconductors is primarily due to non-thermal sources of quasiparticle generation and not slow recombination.

\end{abstract}

\maketitle

\vspace{-0.5cm}

\noindent \textit{Introduction} --- The performance of superconducting quantum circuits depends on a pristine environment to ensure high coherence and reliability. One bottleneck is the presence of broken Cooper pairs known as quasiparticles. It is widely observed that the quasiparticle density in superconducting devices is higher than what is expected from the equilibrium Bardeen-Cooper-Schrieffer (BCS) theory, which predicts an exponential decrease in the density at low temperatures. Excess quasiparticles are a source of decoherence in Josephson-junction based superconducting qubits \cite{Sun_2012, riste_2013, wilen_2021, Vepsalainen_impact_2020}, loss in superconducting resonators \cite{deVisser_2014, Grunhaupt_2018, cardani_reducing_2021}, and noise in kinetic inductance detectors (KIDs) \cite{Day_2003, deVisser_2011, Karatsu_2019}.

The steady-state quasiparticle population is reached when generation mechanisms are in balance with loss processes.  BCS theory predicts the thermal quasiparticle density at finite temperature to be 
\begin{equation}
    n_{qp} (T) = \nu_0 \sqrt{2 \pi k_B T \Delta} \exp{\left (- \Delta / k_b T\right )}
\end{equation}
where $\nu_0$ is the two-spin density of states, $k_B$ is the Boltzmann constant, and $\Delta$ is the superconducting gap energy. At experimentally achievable temperatures of 10 (50) mK, the expected thermal quasiparticle density in aluminum is $\sim 10^{-95} \; \mu \text{m}^{-3}$ ($\sim 10^{-14} \; \mu \text{m}^{-3}$).  In contrast, measured quasiparticle densities in aluminum, using a variety of techniques, are typically 1 -- 10s $\mu \text{m}^{-3}$ \cite{deVisser_2011, deVisser_2014, Temples_2024, Serniak_2018}  and not lower than $\sim10^{-2} \; \mu \text{m}^{-3}$  \cite{riste_2013, Saira_2012, Vepsalainen_impact_2020, Connolly_2024}.

One possible explanation for the excess quasiparticles is pair-breaking by photons, phonons, or particles with energy larger than $2\Delta$ that are not in thermal equilibrium with the superconductor. Examples include photons from warmer stages of the experimental apparatus \cite{Corcoles_2011, Barends_2011, Gordon_2022, Liu_2024, Petersen_2024} and phonons produced in the underlying substrate by background radiation \cite{Vepsalainen_impact_2020, McEwen_resolving_2022, Gusenkova_2022, Fowler_2024, Loer_2024}.  However, observations of the quasiparticle excess persist despite a range of mitigation efforts such as careful filtering of electrical connections to the environment, and known pair-breaking mechanisms such as background radiation have not been shown to be sufficient to explain the excess.  It is therefore interesting to consider other explanations for excess quasiparticles.

An alternative explanation is that quasiparticle loss rates are slower than the predictions of BCS theory.  The dominant loss process, recombination, occurs when two quasiparticles form a Cooper pair with the simultaneous emission of a phonon. 
Bespalov \textit{et al.} \cite{Bespalov_2016} predict that recombination can be exponentially suppressed at low quasiparticle densities when quasiparticles are trapped in sub-gap states, which arise due to impurities and disorder in real superconducting films \cite{Larkin_1972, Balatsky_2006, Feigelman_2012, deGraaf_2020}. Ref. \cite{Bespalov_2016} further hypothesizes that this slow-down in recombination along with the non-equilibrium generation of quasiparticles, for example by background radiation, is sufficient to explain the quasiparticle excess.

In this paper, we extract the key parameters needed to evaluate the theory of \cite{Bespalov_2016} and determine the feasibility of experimentally observing slow recombination in two commonly used ordered superconductors: elemental aluminum and niobium. We analyze current-voltage curves of superconducting tunnel junctions and extract the gap smearing, as described by Abrikosov -- Gor'kov theory \cite{AG_1960}, and an upper bound for the Lifshitz tail of localized states \cite{Larkin_1972} which, to the best of our knowledge, have not been characterized in aluminum and niobium. Using these parameters, we calculate the range of quasiparticle densities where slow-recombination theory holds and find that the onset of slow recombination is predicted to occur at densities lower than those observed in various aluminum and niobium devices.

\vspace{0.05cm}
\noindent \textit{Theory of slow recombination} --- In BCS theory, we have a gapped superconductor where quasiparticle excitations cannot exist inside the gap. Quasiparticles generated by a pair-breaking photon initially have an energy larger than the gap, but they relax back down via phonon emission. Since there are no available states below the gap, they recombine to maintain a steady state population. The recombination rate, $\Gamma_{BCS}$, is proportional to the quasiparticle density, $n_{qp}$ \cite{Kaplan_1976},
\begin{equation}
    \Gamma_{BCS}(n_{qp}) = \bar{\Gamma} n_{qp}
    \label{eq:BCS_recomb}
\end{equation}
where $\bar{\Gamma}$ is a material-specific recombination constant. 

The theory of slow recombination presented in \cite{Bespalov_2016} considers spatial gap fluctuations caused by imperfections and disorder present in real materials that can generate a tail of states below the gap whose density decays exponentially with energy relative to the gap edge. Now quasiparticles relax back down to the gap edge via phonon emission but have available states below the gap to settle into. When a quasiparticle populates a sub-gap state, it is spatially localized and its decay terminates at an energy $\varepsilon_c$ below the gap edge. This energy depends on the gap smearing and characteristic energy decay of the Lifshitz tail which we extract from density of states fits. From $\varepsilon_c$, a localization radius $r_c$ is defined that sets the range of quasiparticle densities where slow recombination theory holds (see Appendices A and B). 

The recombination of these localized quasiparticles depends on the spatial overlap of their wavefunctions. Considering the average distance between quasiparticles, $r$, there are two limits: (1) the low quasiparticle density limit where $r \gg r_c$, so quasiparticles are further away from each other and recombination is exponentially suppressed, and (2) the high quasiparticle density limit where $r \ll r_c$, so the quasiparticles are spatially close to each other, and we recover the BCS result.

The slow recombination rate as a function of quasiparticle density, $n_{qp}$ can be expressed as \cite{Bespalov_2016},
\begin{equation}
    \Gamma_{B}(n_{qp}) = D \bar{\Gamma} r_c^{-3} \left ( \frac{3 C_p}{4 \pi r_c^3~ n_{qp}} \right) ^{\beta / 3} e^{-\left ( \frac{3 C_p}{4 \pi r_c^3 ~ n_{qp}} \right)^{1/3}}
\label{eq:Besp_recomb}
\end{equation}
where $D \approx 0.06$, $\beta = 0.41$, and $C_p = 0.605$. The critical quasiparticle density, $n_{qp,c} = 5 \times 10^{-3} \; r_c^{-3}$, sets the density range for which slow recombination holds. Therefore, recombination is exponentially suppressed when $n_{qp} < n_{qp,c}$, while for $n_{qp} > n_{qp,c}$, the BCS result is recovered.

\begin{figure}[ht!]
    \centering
    \includegraphics[scale = 0.49]{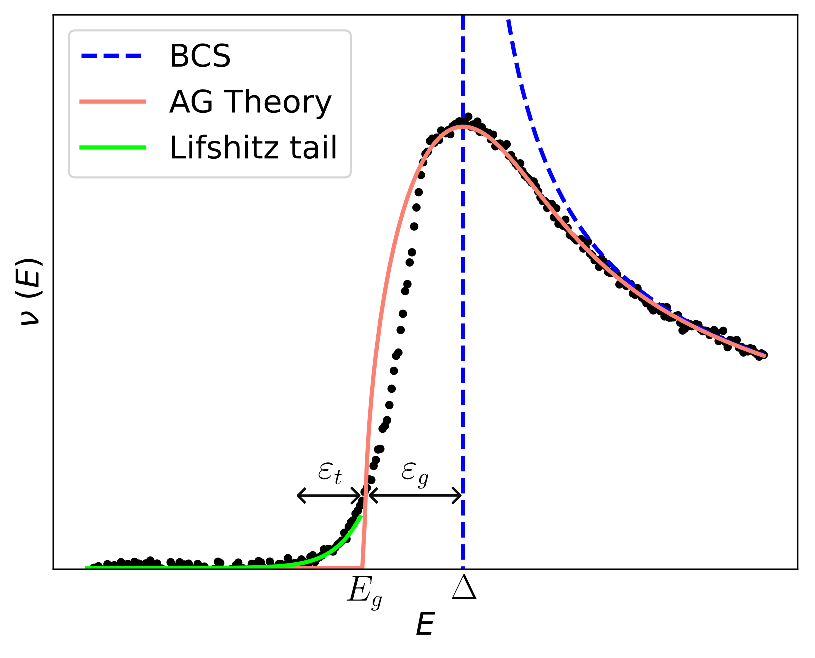}
    \vspace{-0.45cm}
    \caption{Black points represent a typical density of states measurement. Dashed blue curve is the calculated BCS density of states with a hard gap at $E$ = $\Delta$. Non-idealities in the DOS can be expressed as a broadening of the BCS singularity and sub-gap states below the shifted hard gap at $E = E_g$, that is quantified by Abrikosov and Gor'kov \cite{AG_1960} theory (salmon curve) with energy $\varepsilon_g$, and Lifshitz tail \cite{Larkin_1972} (green curve) with a characteristic energy decay $\varepsilon_t$, respectively.}
    \label{fig1}
     \vspace{-0.25cm}
\end{figure}

\vspace{0.05 cm}
\noindent \textit{Density of states fitting} ---  To evaluate the recombination rate of Eq. \ref{eq:Besp_recomb}, we carefully characterize deviations from the BCS density of states. As we can see in Fig. \ref{fig1}, in real materials there is both a rounding of the BCS singularity and sub-gap states, captured by the energy scales $\varepsilon_g$ and $\varepsilon_t$, respectively. Here we outline our fitting routine which combines Abrikosov-Gor'kov theory \cite{AG_1960} to capture the rounding of the BCS singularity with a Lifshitz tail \cite{Larkin_1972} to capture the sub-gap states.
%\vspace{-0.25cm}

\begin{figure*}[ht!]
    \begin{subfigure}[b]{0.5\textwidth}
        %\centering
        %\caption{}
        \vspace{-0.2cm}
        \includegraphics[scale = 0.5]{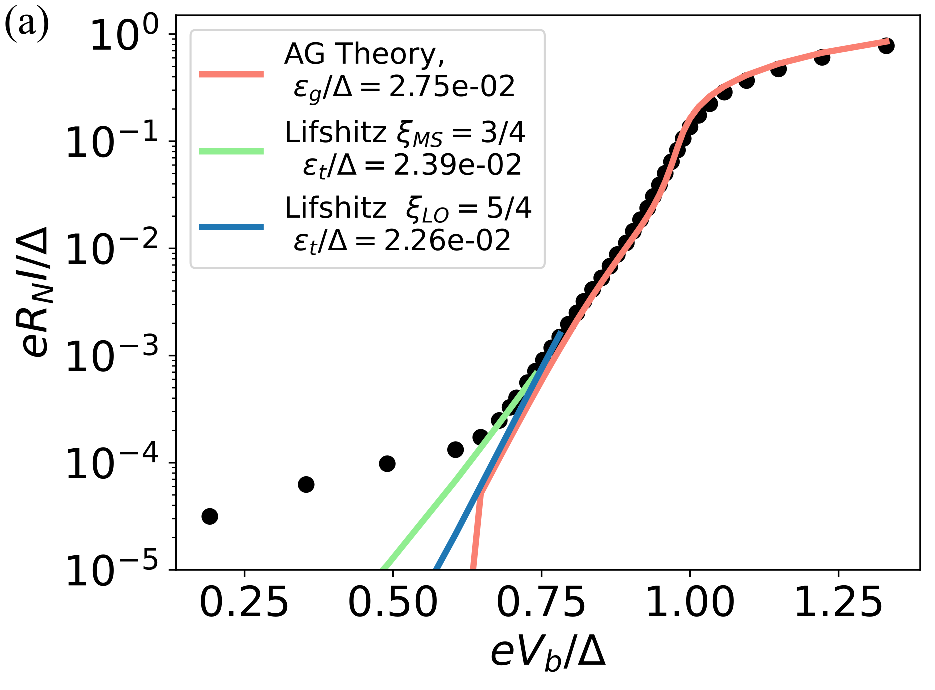}
        
    \end{subfigure}%
    ~
    \begin{subfigure}[b]{0.5\textwidth}
        %\centering
        %\caption{}
        \vspace{-0.2cm}
        \includegraphics[scale = 0.5]{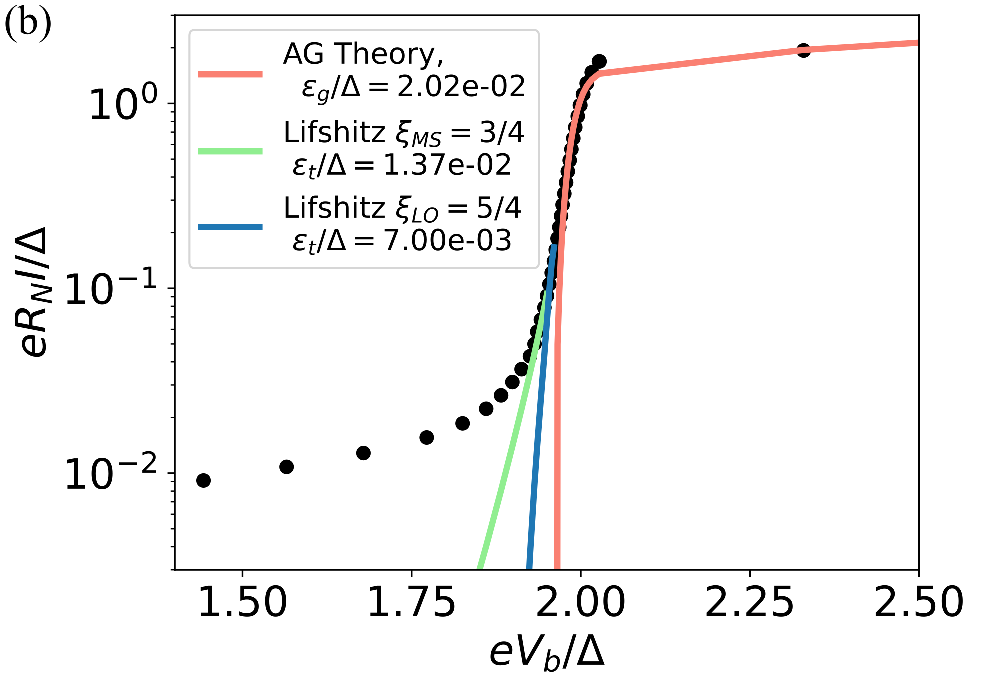}
        
    \end{subfigure}
    \vspace{-0.6cm}
    \caption{
    Measured current-voltage curves for superconducting tunnel junctions (black points) and fits using AG theory (salmon curve) to extract $\varepsilon_g/\Delta$ and two Lifshitz tails (formulation is given in Appendix B) to extract $\varepsilon_t/\Delta$ for exponents $\xi_{MS}$ (green curve) and $\xi_{LO}$ (blue curve). (a) Data and fits for an Al \textbf{--} AlOx \textbf{--} AlMn NIS junction at 89 mK. Salmon curve includes a voltage-dependent temperature calculated using a model of NIS junction self-cooling \cite{ONeil_2012}. (b) Data and fits for a Nb \textbf{--} AlOx \textbf{--} Nb SIS junction at 67 mK.}
    \label{fig2}
     \vspace{-0.5cm}
\end{figure*}

Non-idealities in the BCS density of states (DOS) have been theoretically studied and observed. Abrikosov and Gor'kov (AG) described the effect of magnetic impurities by a pair-breaking parameter $\eta$ \cite{AG_1960}. For small values of $\eta$, the DOS retains a hard gap but the gap edge shifts to $E_g$ due to the rounding of the BCS singularity as can be observed in Fig~\ref{fig1}. The scale of the rounding is $\varepsilon_g = \Delta - E_g$.   Details of the AG DOS formulation are given in Appendix A.

Larkin and Ovchinnikov \cite{Larkin_1972} later showed that the problem of magnetic impurities is the same as considering fluctuations in the pairing potential. At the mean field level the AG result is recovered. Beyond mean field, rare optimal fluctuations (for example, in the presence of disorder) result in sub-gap states described by a Lifshitz tail. The characteristic decay energy of the tail is $\varepsilon_t$ as shown in Fig~\ref{fig1}. The terminal quasiparticle energy $\varepsilon_c$ is related to $\varepsilon_t$. Here we make a step towards identifying the effects of these fluctuations in our data.

To extract the characteristic decay of the Lifshitz tail, we use a stretched exponential of the form,
\begin{equation}
    \nu_{L}(E) \propto \exp{ \left [ - \left ( \frac{E_g - E}{\varepsilon_t} \right )^{\xi} \right ]}
    \label{eq:Lifshitz}
\end{equation}
where $\xi$ is the characteristic exponent. Larkin and Ovchinnikov \cite{Larkin_1972} derived an exponent $\xi_{LO}$ by studying optimal fluctuations of $\Delta$ in the Usadel equation close to the mean-field gap edge. Details on the functional form of the Lifshitz tail and the resulting $\varepsilon_c$ and $r_c$ parameters for both the original formulation in \cite{Bespalov_2016} with $\xi_{LO}$ and the consequences of an alternative formulation of the Lifshitz tail with $\xi_{MS}$ given in \cite{Meyer_2001, meyer_thesis} are described in Appendix B.

Thus, we have a DOS which we can express as,
\begin{equation}
  \nu(E) =  
   \begin{dcases}
        \nu_{AG}(E, \Delta, \eta),  & E \ge E_g  \\
        \nu_{L}(E , E_g, \varepsilon_t ),  & E < E_g
    \end{dcases}
    \label{eq:FullDOS}
\end{equation}
In this model, there are three free parameters, $\Delta$, $\eta$, and $\varepsilon_t$. Parameter $\varepsilon_g$ is constrained by $\Delta$ and $\eta$ (see Appendix A).

\begin{table*}[ht!]
\centering
    \caption{Summary of key fit parameters ($\varepsilon_g / \Delta$, $\varepsilon_{t,LO} / \Delta$) extracted from measured aluminum and niobium devices and lower Dynes parameter devices reported in the literature, calculated localization radii $r_{c,LO}$, and critical quasiparticle density $n_{qp,c}$. }
    \vspace{-0.25cm}
    \begin{tabularx}{1.0 \textwidth}{| >{\centering\arraybackslash}X 
  | >{\centering\arraybackslash}X 
  | >{\centering\arraybackslash}X
  | >{\centering\arraybackslash}X
  | >{\centering\arraybackslash}X |}
    \hline
        Device & $\varepsilon_g / \Delta$ & $\varepsilon_{t,LO} / \Delta$ & $r_{c,LO}$ (nm) & $n_{qp,c} \; (\mu \text{m}^{-3})$\\ 
        \hline \hline
        Al NIS & $2.75 \times 10^{-2}$ & $2.26 \times 10^{-2}$ & 553 & $2.9 \times 10^{-2}$ \\ 
        \hline
        Ref. \cite{Saira_2012} & $3.56 \times 10^{-7}$ & $2.47 \times 10^{-7}$  & 2920  &  $2 \times 10^{-4}$ \\
        \hline \hline
        Nb SIS & $2.02 \times 10^{-2}$ & $7 \times 10^{-3}$& 121 & 3 \\
        \hline
        Ref. \cite{Hatinen_2024} & $1.28 \times 10^{-3}$ &$1.18 \times 10^{-3}$ & 188 & 0.75 \\
         \hline
  
  \end{tabularx}
    
    \label{tab1}
\end{table*}

\begin{figure*}[ht!]
    \begin{subfigure}[b]{0.5\textwidth}
        \centering
        %\caption{}
        \vspace{-0.2cm}
        \includegraphics[scale = 0.5]{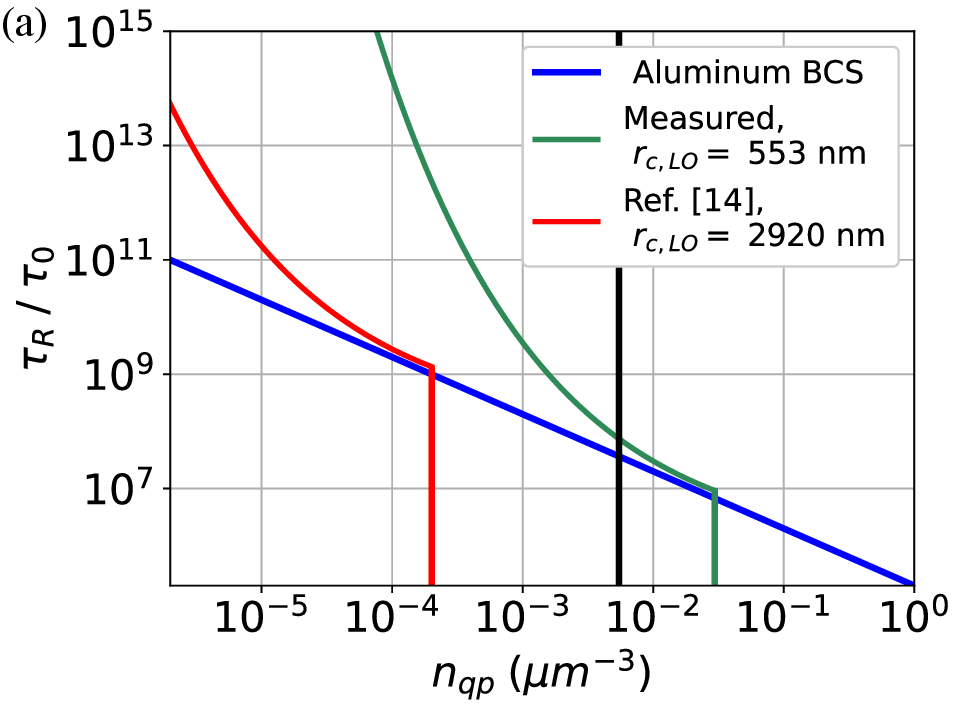}
    \end{subfigure}%
    ~ 
    \begin{subfigure}[b]{0.5\textwidth}
        \centering
        %\caption{}
        \vspace{-0.2cm}
        \includegraphics[scale = 0.5]{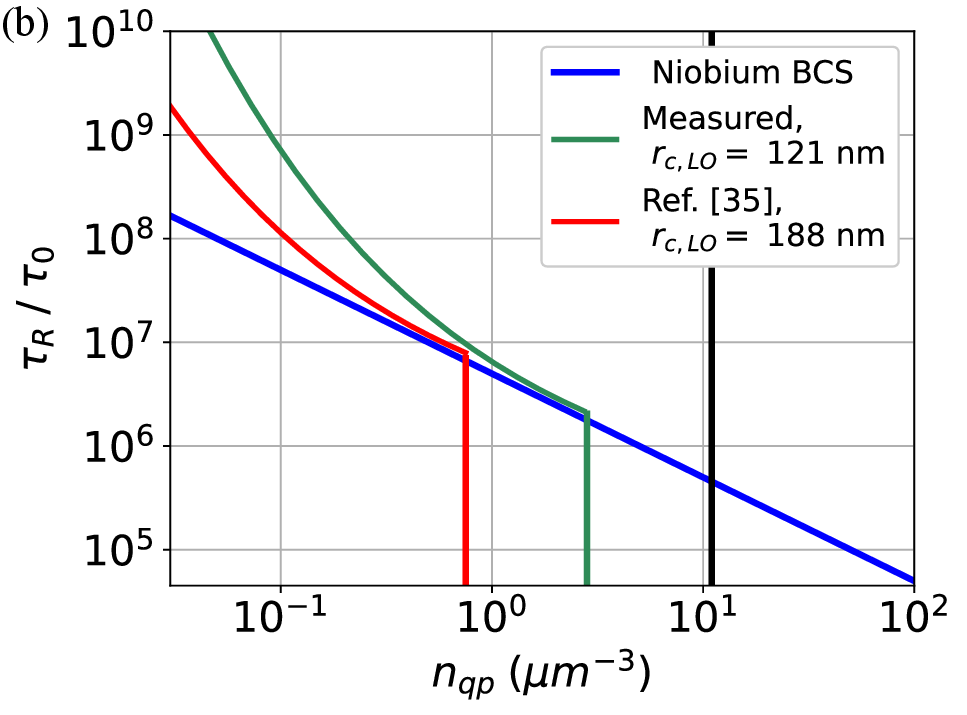}
    \end{subfigure}
     \vspace{-0.6cm}
    \caption{Recombination time $\tau_R$ in units of the electron-phonon coupling time $\tau_0$ \cite{Kaplan_1976} plotted as a function of $n_{qp}$ for aluminum (left) and niobium (right). Blue curves correspond to the BCS recombination time $\tau_R = 1 / \Gamma_{BCS}$ (Eq. \ref{eq:BCS_recomb}). Red and green curves correspond to recombination time $\tau_R = 1 / \Gamma_{B}$ (Eq. \ref{eq:Besp_recomb}) from literature and our measurements, respectively. Red and green vertical lines denote $n_{qp,c}$ values from the literature and our measurements, respectively. Vertical black lines denote the lowest measured $n_{qp}$ in the literature. (a) Predicted recombination times in aluminum. Our measurements yield a higher $n_{qp,c} = 2.9 \times 10^{-2}~\mu \text{m}^{-3}$ and Ref. \cite{Saira_2012} yields a lower $n_{qp,c} = 2 \times 10^{-4}~\mu \text{m}^{-3}$. The lowest measured $n_{qp}$ in the literature \cite{Connolly_2024} is $5 \times 10^{-3}\mu \text{m}^{-3}$. (b) Predicted recombination times in niobium. Our measurements yield a higher $n_{qp,c} = 3~ \mu \text{m}^{-3}$ and Ref. \cite{Hatinen_2024} yields a lower $n_{qp,c} = 0.75 ~ \mu \text{m}^{-3}$. The lowest measured $n_{qp}$ inferred from the literature is $\sim 11 \;\mu \text{m}^{-3}$ \cite{Dominjon2019} .}
    
    \label{fig3}
     \vspace{-0.25cm}
\end{figure*}

\vspace{0.06cm}

\noindent \textit{Results} --- To probe the DOS of aluminum and niobium, we take current-voltage measurements of tunnel junctions: (1) AlMn -- AlOx -- Al normal metal-insulator-superconductor (NIS) junctions and (2) Nb -- AlOx -- Nb superconductor-insulator-superconductor (SIS) junctions. The current through the junction as a function of applied voltage can be written as,
\begin{equation}
\begin{split}
    I(V_b) = \frac{1}{e R_N} \int_{-\infty}^{\infty} \text{d} E \; \nu_{l}(E) \; \nu_{r}(E - e V_b) \; \times  \\
    \left ( f_r (E - e V_b) - f_l (E) \right)
\end{split}
\end{equation}
\noindent where $e$ is the electron charge, $R_N$ is the normal resistance of the junction, $\nu_{l}$, $\nu_{r}$ are the DOS of the left and right junction electrodes, respectively, and $f(E) = \left (e^{E/ k_b T} + 1 \right )^{-1}$ is the Fermi function. The details of the measurements and a summary of all the measured devices are described in the Supplemental Material \cite{SI_ref}.

Representative fits to the current-voltage data using the DOS in Eq. \ref{eq:FullDOS} are shown in Fig \ref{fig2}. We observe that AG theory captures our data near $\Delta$ but diverges from the data in the sub-gap. Including the Lifshitz tail improves the fit of the sub-gap data but does not capture points in the deep sub-gap where a linear leakage current persists.  This leakage current is commonly observed in tunnel junctions at low voltages both in measurements of aluminum and niobium NIS junctions \cite{Pekola_2010, Saira_2012, groll_2018, Bafia_2022, Hatinen_2024} and niobium tri-layer SIS junctions \cite{Noguchi_2011, Tolpygo_2017}. By choice, our DOS model does not include a separate term to capture this deep sub-gap leakage to ensure that $\varepsilon_t$ is an upper bound on this quantity.
 
A variety of phenomena can cause sub-gap leakage including junction barrier pinholes and flux trapping. In addition, photon-assisted tunneling causes both sub-gap leakage and gap smearing \cite{Pekola_2010}. The Dynes parameter $\Gamma$ is commonly used to describe both leakage current and gap smearing \cite{Dynes_1984}. Fits of our measured current-voltage curves using the Dynes model are shown in the Supplemental Material \cite{SI_ref} where we find Dynes parameters similar to many literature values. Because of the various contributions to sub-gap leakage and gap smearing that are unrelated to the true density of states of the superconductors, our fits for $\varepsilon_g$ and $\varepsilon_t$, represent upper bounds on these energies.

We can make the upper bounds on $\varepsilon_g$ and $\varepsilon_t$ more realistic by considering measurements in the literature that achieved lower Dynes parameters than we found in our devices. These lower Dynes parameters can be attributed to better shielding, filtering, and junction quality, and it can be argued that they provide a more accurate representation of the true density of states. In particular, for aluminum NIS junctions, Saira et al. \cite{Saira_2012}, and for niobium NIS junctions, Hätinen et al.  \cite{Hatinen_2024}, achieved record-low Dynes parameters. We predict the associated $\varepsilon_g$ and $\varepsilon_t$ values for these devices using their published Dynes parameters and techniques given Appendix C. A summary of the fit parameters $\varepsilon_g$ and $\varepsilon_{t,LO}$, and calculated $r_c$ and $n_{qp,c}$ values are given in Table \ref{tab1} for our aluminum and niobium measurements and for the more ideal aluminum and niobium measurements in the literature. 

\vspace{0.05cm}
\noindent \textit{Discussion} --- The parameters extracted from fitting the current-voltage curves give an upper bound on gap smearing, $\varepsilon_g$, and Lifshitz tail decay, $\varepsilon_t$. Using these to calculate localization radii $r_c$, we can predict an upper bound on the critical quasiparticle density $n_{qp,c}$ that marks the onset of the exponential slowdown in recombination.

We plot the predictions of slow recombination theory and BCS theory in Fig. \ref{fig3}. The red and green curves are calculated using $\varepsilon_g$ and $\varepsilon_t$ from the more ideal junction measurements in the literature \cite{Saira_2012, Hatinen_2024}, and from our own tunnel junction measurements, respectively.  The $n_{qp,c}$ values predicted from these localization radii give a realistic upper bound on the range of cut-off quasiparticle densities for the onset of slow recombination. We assess the validity of the slow recombination theory by comparing these $n_{qp,c}$ values to both the lowest measured densities in the literature and to more typical measured values.

For aluminum, we can see in Fig. \ref{fig3}a that the highest $n_{qp,c} = 2.9 \times 10^{-2} ~\rm{\mu m^{-3}}$, calculated for our measured devices, is smaller than typical densities  measured in qubits which fall between $\sim 0.04 - 0.1 ~\rm{\mu m^{-3}}$ \cite{riste_2013, Vepsalainen_impact_2020, Serniak_2018} and $\sim 3$ orders of magnitude smaller than the lowest densities observed in aluminum resonators which are about $\sim 10~\rm{\mu m^{-3}}$ \cite{deVisser_2011, deVisser_2014, Temples_2024}. The very lowest measured quasiparticle density of $\sim 5 \times 10^{-3} ~\rm{\mu m^{-3}}$ \cite{Connolly_2024} falls  below the highest $n_{qp,c}$ predicted for our measured devices but is still over an order of magnitude larger than the lower $n_{qp,c} = 2 \times 10^{-4}~\rm{\mu m^{-3}}$, calculated for the more ideal literature DOS measurements. Since the lowest measured quasiparticle density is an order of magnitude higher than the most realistic upper bound for the onset of slow recombination, and since typical quasiparticle densities in qubits and resonators are larger than the $n_{qp,c}$ derived from our own devices, we conclude that slow recombination is not the cause of excess quasiparticle densities observed to date in aluminum.

Observed quasiparticle densities in niobium vary widely but consistently exceed the cut-off densities predicted by both the green and red curves in Fig. \ref{fig3}b. A recent result in a niobium-based qubit device \cite{Anferov_2024} shows qubit lifetime saturation for Nb at $\approx$ 1.1 K which corresponds to a QP density of $33~\rm{\mu m^{-3}}$, higher than the $n_{qp,c}$ for both the red and green curves in Fig. \ref{fig3}b. This is consistent with the temperatures where the internal quality factor saturates in MKIDs, which serves as another probe of the excess quasiparticle density. For typical high-quality niobium films, with critical temperatures $T_c$ ranging from 9.2 - 9.5 K, and high-quality factor niobium SRF cavities \cite{Romanenko_2020}, the quality factor saturation occurs between $\sim 0.1~T_c-0.2~T_c$ yielding quasiparticle densities as low as $11 - 17~\rm{\mu m^{-3}}$ \cite{Dominjon2019, Zhu_2022} and as high as $10^{3} - 10^{4}~\rm{\mu m^{-3}}$ \cite{Barends_2007, Noguchi_2019}. The lowest measured density of $11 ~\rm{\mu m^{-3}}$ \cite{Dominjon2019} is comparable to our highest $n_{qp,c} = 3~\rm{\mu m^{-3}}$, calculated for our measured devices, and is about one order of magnitude higher than the lower $n_{qp,c} = 0.75 ~\rm{\mu m^{-3}}$, calculated from the most ideal DOS measurements in the literature. Therefore, we conclude that slow recombination is not the cause of the excess densities in niobium.

\vspace{0.017cm}
\noindent \textit{Conclusion} --- We have evaluated the theory of slow quasiparticle recombination \cite{Bespalov_2016} for ordered aluminum and niobium. The theory hypothesizes that the quasiparticle excess universally observed in superconducting films is caused by slow recombination and non-equilibrium generation. Quantitative evaluation of the slowdown requires detailed knowledge of the superconducting density-of-states. Using our own density-of-states measurements from superconducting tunnel junctions as well as measurements from the literature, we have extracted values for the AG gap smearing $\varepsilon_g$ and the sub-gap Lifshitz tail $\varepsilon_t$ for ordered films of aluminum and niobium. With these parameters, we calculate critical quasiparticle densities below which slow recombination should be present. We find that the predicted onset of slow recombination occurs at quasiparticle densities below the densities observed in superconducting films. Our results thus show that slow recombination is not responsible for the quasiparticle excess in commonly used ordered superconductors. 

Future experimental work in this area should look at recombination in disordered superconductors where slow recombination is predicted to occur at higher quasiparticle densities and thus may be more easily visible. Presently, measurements of recombination in disordered materials are inconclusive \cite{Grunhaupt_2018, Barends2009, Coumou2012}. Future work should also evaluate whether scattering by thermal phonons or readout photons can excite quasiparticles out of the Lifshitz tail, thereby reducing localization effects that are hypothesized to become more prominent as disorder increases.

\noindent \textit{Acknowledgements} --- We thank NIST ERB reviewers Jordan Wheeler and John Teufel for valuable feedback on the manuscript. We thank the U.S. Army Research Office for financial support through the Quantum Computing in the Solid State with Spin and Superconducting Systems (QC-S$^5$) program. E.G. acknowledges financial assistance award 70NANB23H027 from  the U.S. Department of Commerce, National Institute of Standards and Technology.

\smaller[19]
\bibliography{references.bib}

\vspace{20cm}
\newpage

\normalsize
\onecolumngrid
\section*{End Matter}

\twocolumngrid

\noindent \textit{Appendix A: AG theory DOS formulation} --- Here, we provide the functional form of the AG \cite{AG_1960} density of states used to fit data in Fig. \ref{fig2} of the main text and Fig. \ref{fig5} in Appendix C. To model the rounding of the BCS singularity, we start with the mean-field treatment of the Usadel equation \cite{Larkin_1972, Feigelman_2012}, 
\begin{equation}
    iE \sin{\theta_{AG}} + \bar{\Delta} \cos{\theta_{AG}} - \bar{\Delta} \eta \sin{\theta_{AG}} \cos{\theta_{AG}} = 0
\end{equation}
where $\eta$, is an effective depairing parameter, $\theta_{AG}$ is the pairing angle, and $\bar{\Delta}$, is the average value of the superconducting gap. Using this equation to solve for $\theta_{AG}$, the gapped quasiparticle density of states can be expressed as,
\begin{equation}
    \nu_{AG}(E) = \rm{Re} (\cos{\theta_{AG}})
\end{equation}
In this case, the BCS coherence peak is smeared and the hard gap moves to $E_g$ \cite{Meyer_2001, Skvortsov_subgap_2013},
\begin{equation}
    E_g = \bar{\Delta}(1 - \eta^{2/3})^{3/2}
\end{equation}
From this, we obtain $\varepsilon_g = \bar{\Delta} - E_g$. In the limit of $\eta \rightarrow 0$, we recover the BCS density of states.

\noindent \textit{Appendix B: Lifshitz tail DOS formulation} --- Larkin and Ovchinnikov \cite{Larkin_1972} obtained an exponent $\xi_{LO} = (8 - d)/4$ by looking at optimal fluctuations of $\Delta$ in the Usadel equation close to the mean-field gap edge. The problem was revisited by Meyer and Simon \cite{meyer_thesis, Meyer_2001}, where further studies of the nonlinearity of the Usadel equation were done and an exponent $\xi_{MS} = (6 - d) /4$ was derived. This description is applicable for energies further from the gap-edge \cite{Skvortsov_subgap_2013}. Ref \cite{Bespalov_2016} derives the localization radius for the Lifshitz tail with exponent $\xi_{LO} = 5/4$, and here we also consider an extension of those results for the Lifshitz tail with exponent $\xi_{MS} = 3/4$. These two localization radii gives a range of critical densities $n_{qp,c}$ that mark the onset of slow recombination theory.

Starting with the formulation used in Ref. \cite{Bespalov_2016}, the Lifshitz tail for a 3D film has an exponent $\xi_{LO} = 5/4$ and can be expressed as:
\begin{equation}
    \nu_{LO}(\varepsilon) = A \nu_0  (\varepsilon/\varepsilon_t )^{9/8}  \exp\left [-(\varepsilon/\varepsilon_t)^{5/4}  \right]
\end{equation}
where $A = 0.79 \frac{\sqrt{\varepsilon_t/\Delta}}{\varepsilon_g/\Delta}$ and $\varepsilon = E_g - E$ is the distance from the gap edge. To compute the number of localized states that overlap with a given state, we integrate over the Lifshitz tail:
\begin{equation}
\begin{split}
    N(\varepsilon) = L^{3}(\varepsilon) \int_{\varepsilon}^{\infty} d\varepsilon' \nu_L(\varepsilon') \\ \approx 
    N_T \left (\varepsilon/\varepsilon_t\right)^{1/8} \exp \left [- (\varepsilon/\varepsilon_t)^{5/4}  \right]
\end{split}  
\end{equation}
where  $L(\varepsilon) = (2/3)^{1/4} \xi \left( \varepsilon / \Delta \right)^{-1/4}$ and $N_T = (4/5)A \nu_0 \varepsilon_t L(\varepsilon_T)^3 $. The coherence length $\xi$ is calculated in the dirty limit as $\xi \approx (\xi_{BCS}t)^{1/2}$ for a film of thickness $t$. $N_T$ represents the number of localized states at 
$\varepsilon \simeq \varepsilon_t$ and can be approximated as $N_{T} \simeq N(\varepsilon_t) \sim g \frac{(\varepsilon_t/\Delta)^{3/4}}{ \varepsilon_g/\Delta}$ where $g = \pi \nu_0 \Delta \xi^3$. Localized quasiparticles end their relaxation at the terminal energy, $\varepsilon_c$, defined when $N(\varepsilon_c) \simeq 1 $ which is approximately given by:
\begin{equation*}
    \varepsilon_{c,LO} \approx \varepsilon_{t,LO} (\ln N_{T})^{4/5}
\end{equation*}
The localization radius is defined as $r_{c,LO} = L(\varepsilon_{c,LO}) / 2$. In the main text, we give the values of these parameters in Table \ref{tab1} and plot the predicted slow recombination time as a function of quasiparticle density in Fig. \ref{fig3} for these parameters.

An alternative formulation based on Ref. \cite{Meyer_2001, meyer_thesis} yields a Lifshitz tail for a 3D film with an exponent $\xi_{MS} = 3/4$ and can be expressed as:
\begin{equation}
    \nu_L(\varepsilon) = A \nu_0 \cdot \exp\left [-\left (\varepsilon/\varepsilon_t \right)^{3/4}  \right]
    \label{eq:Lifshitz_MS}
\end{equation}
Following the same procedure as done with $\xi_{LO}$, we numerically compute the number of localized states that overlap with a given state but integrating over the Lifshitz tail in eq. \ref{eq:Lifshitz_MS} this time.

\begin{figure}[ht!]
    \centering
    \includegraphics[scale = 0.44]{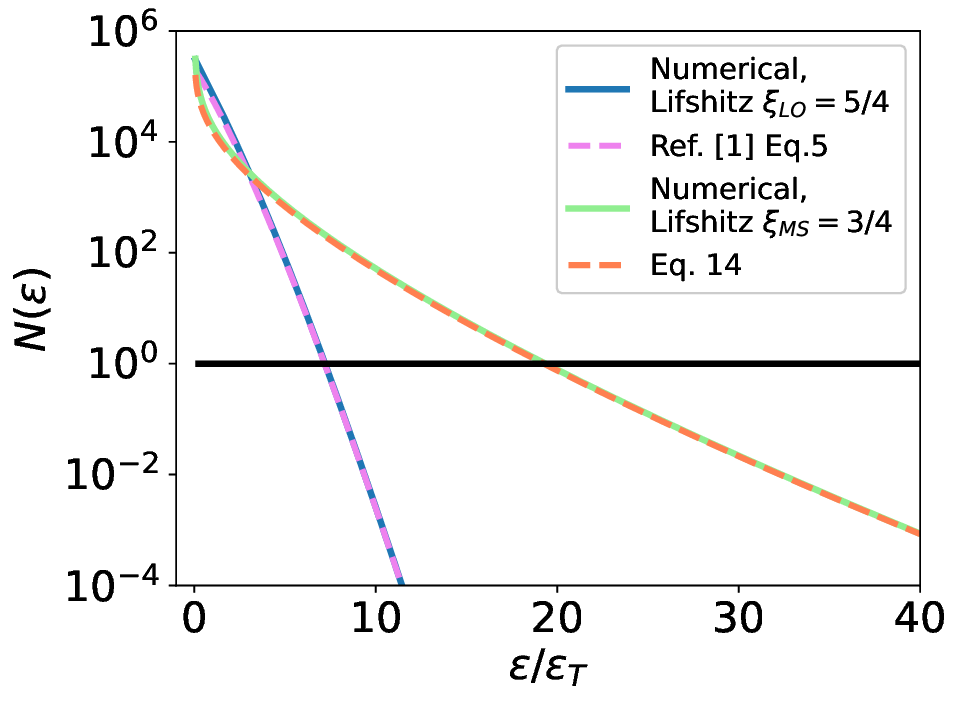}
    \caption{Comparison of the numerical solution of the number of localized states that overlap with a given state, $N(\varepsilon)$, to the approximate solutions. For $\xi_{LO} = 5/4$, the numerical solution (blue) is compared to the approximate solution given in Eq. 5 of \cite{Bespalov_2016} or Eq. 11 in this paper (magenta). For $\xi_{MS} = 3/4$, the numerical solution (green) is compared to the approximate solution given in Eq. 14 (orange). The black line marks $N(\varepsilon) = 1$ and the energy at which the curves intersect with this line define $\varepsilon_c$.}
    \label{fig:Numerical_N1}
\end{figure}
\begin{table*}[ht!]
\centering
    \caption{Summary of key fit parameters ($\varepsilon_g / \Delta$, $\varepsilon_{t,MS} / \Delta$), extracted from measured aluminum and niobium devices and lower Dynes parameter devices reported in the literature, calculated localization radii $r_{c,MS}$, and critical quasiparticle density $n_{qp,c}$. This table replicates Table \ref{tab1} except for the use of eq. \ref{eq:Lifshitz_MS} to describe the Lifshitz tail.} 
    \vspace{-0.25cm}
    \begin{tabularx}{1.0 \textwidth}{| >{\centering\arraybackslash}X 
  | >{\centering\arraybackslash}X 
  | >{\centering\arraybackslash}X
  | >{\centering\arraybackslash}X
  | >{\centering\arraybackslash}X |}
    \hline
        Device & $\varepsilon_g / \Delta$ & $\varepsilon_{t,MS} / \Delta$ & $r_{c,MS}$ (nm) & $n_{qp,c} \; (\mu \text{m}^{-3})$\\ 
        \hline \hline
        Al NIS & $2.75 \times 10^{-2}$ & $2.39 \times 10^{-2}$ & 409 & $7.3 \times 10^{-2}$ \\ 
        \hline
        Ref. \cite{Saira_2012} & $3.56 \times 10^{-7}$ & $3.31 \times 10^{-7}$  & 2510  &  $3.2 \times 10^{-4}$ \\
        \hline \hline
        Nb SIS & $2.02 \times 10^{-2}$ & $1.37 \times 10^{-2}$& 79 & 10 \\
        \hline
        Ref. \cite{Hatinen_2024} & $1.28 \times 10^{-3}$ &$1.22 \times 10^{-3}$ & 150 & 1.5 \\
         \hline
  
  \end{tabularx}
    
    \label{tab2}
    %\vspace{-3cm}
\end{table*}

\begin{figure*}[ht!]
\twocolumngrid
    %\vspace{-5cm}
    \begin{subfigure}[h]{0.5\textwidth}
        \centering
        %\caption{}
        \vspace{-0.2cm}
        \includegraphics[scale = 0.46]{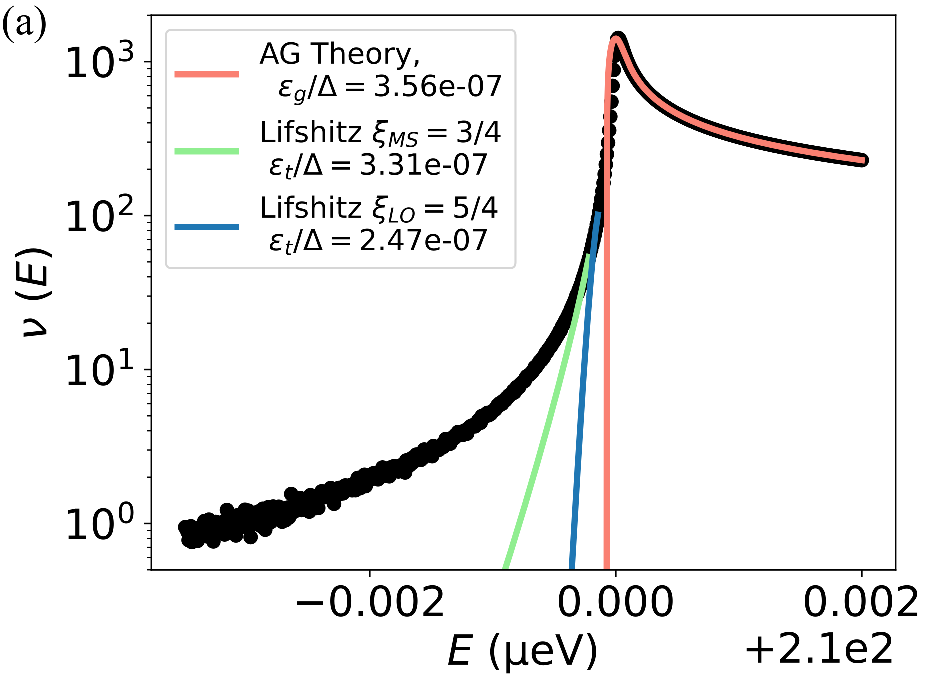}
    \end{subfigure}% 
    \hfill
    \begin{subfigure}[h]{0.5\textwidth}
        \centering
        %\caption{}
        \vspace{-0.2cm}
        \includegraphics[scale = 0.46]{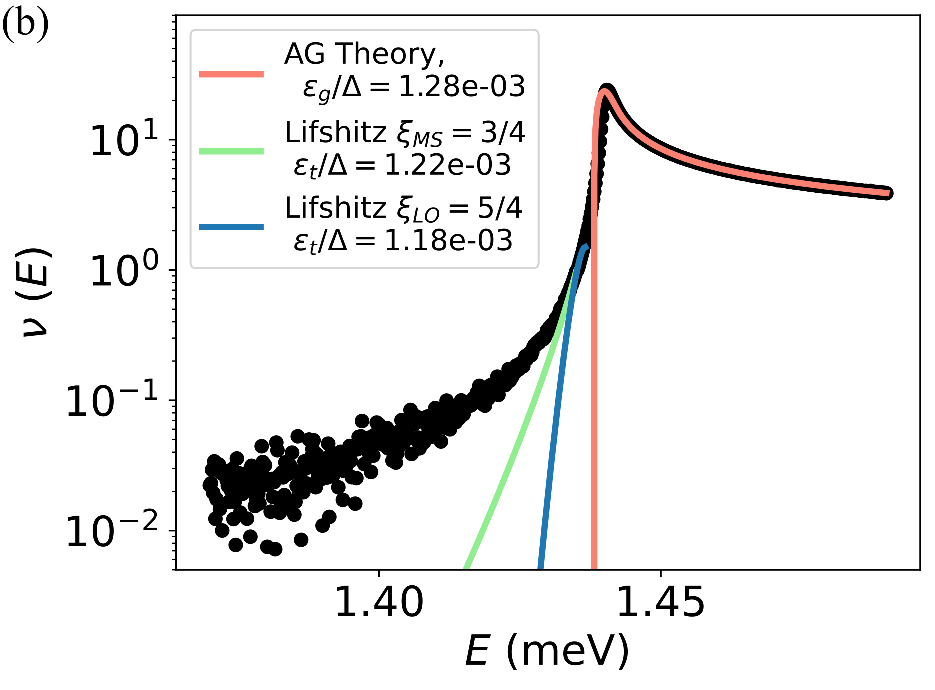 }
    \end{subfigure}
    \caption{Representative DOS fits to simulated data in black with AG theory (shown by the solid salmon color line) and the Lifshitz tail (shown by the solid green color line). (a). The simulated data corresponds to one of the lowest Dynes parameters measured in literature for aluminum in Saira et al. of $\Gamma / \Delta = 1.7 \times 10^{-7}$ \cite{Saira_2012}. The coherence length in the dirty limit is $\xi \approx 245 $ nm . (b) The simulated data corresponds to a Dynes parameter of $\Gamma / \Delta = 5.6 \times 10^{-4}$ measured in Hätinen et al. \cite{Hatinen_2024} for niobium. The coherence length in the dirty limit is $\xi \approx 130 $ nm. }
    \label{fig5}
    %\vspace{-5cm}
\end{figure*}
\noindent Using eq. \ref{eq:Lifshitz_MS}, we can approximate the number of states as: 
\begin{equation}
\begin{split}
        N(\varepsilon) = L^{3}(\varepsilon) \int_{\varepsilon}^{\infty} d\varepsilon' \nu_L(\varepsilon') \\
        \approx N_T \left (\varepsilon/\varepsilon_t\right)^{-1/2} \exp\left [-(\varepsilon/\varepsilon_t)^{3/4}  \right]
\end{split}
\end{equation}
where $N_T = (4/3)A \nu_0 \varepsilon_t L(\varepsilon_T)^3$. A comparison of the numerical results to the approximate results for both $\xi_{MS}$ and $\xi_{LO}$ cases is shown in Fig. \ref{fig:Numerical_N1}. 

\noindent For this formulation, $\varepsilon_c$ is given by the equation:
\begin{equation*}
    \left (\varepsilon_{c,MS}/\varepsilon_{t,MS} \right)^{3/4} = \ln N_T - (1/2) \ln \left (\varepsilon_{c,MS}/\varepsilon_{t,MS} \right)
\end{equation*}
And we can define localization radius $r_{c,MS} = L(\varepsilon_{c,MS}) / 2$.  

Table \ref{tab2} is a summary of the fit parameters using the formulation with $\xi_{MS}$ and the corresponding $r_{c,MS}$ and $n_{qp,c}$. We can see that the $r_{c,MS}$ values are smaller than $r_{c,LO}$ and therefore yield a higher $n_{qp,c}$. These differences can be attributed to the fact that the Lifshitz tail with exponent $\xi_{MS}$ is a better fit to the data deeper in the sub-gap and therefore $\varepsilon_{t,MS}$ is likely inflated by contributions from sub-gap leakage. However, the relative ordering of the $n_{qp,c}$ values in Table \ref{tab2} and the observed quasiparticle densities summarized in the main text is largely unchanged from the $\varepsilon_{t,LO}$ case of Table \ref{tab1}. Thus, we can still conclude that the predicted onset of slow recombination is inconsistent with observed densities in aluminum and niobium devices.

%\vspace{-5cm}
\noindent \textit{Appendix C: DOS fits based on literature data} --- We look at junction measurements in the literature of NIS junctions which have yielded even lower Dynes parameters in aluminum  and niobium than what we have measured. These lower Dynes parameter values can be attributed to better filtering, shielding, and junction quality. For the aluminum measurement, capacitive shunting was employed to suppress sub-gap leakage due to photon-assisted tunneling, therefore yielding a very low Dynes parameter \cite{Saira_2012}. For the niobium junction, the fabrication was carefully tuned to yield junctions with low sub-gap leakage \cite{Hatinen_2024}. Therefore, we can view these measurements as getting closer to representing the true density of states for aluminum and niobium.

\twocolumngrid
Based on the Dynes parameters $\Gamma$ reported in \cite{Saira_2012} and \cite{Hatinen_2024}, we generated simulated DOS curves using the Dynes model which was demonstrated to be a good fit to their data. The Dynes DOS is given by \cite{Dynes_1984},
\begin{equation}
    \nu(E) = \rm{Re} \left ( \frac{E + i \Gamma}{\sqrt{ (E + i \Gamma)^{~2} - \Delta^2}} \right ).
\end{equation}
Using this model, we generate simulated data and fit the DOS using  AG theory and the Lifshitz tails, as described in Appendices A and B, respectively, to estimate $\varepsilon_g$ and $\varepsilon_t$. The fits to the simulated Dynes curves are shown in Fig. \ref{fig5}. These values yield an improved upper bound on $n_{qp,c}$ that better represents ordered aluminum and niobium devices.

\end{document}

\title{Supplemental Material: 
Can slow recombination in ordered superconductors explain the excess quasiparticle population?
}

\author{Eva Gurra}
\email{eva.gurra@colorado.edu}
\affiliation{National Institute of Standards and Technology, Boulder, CO 80305}
\affiliation{Department of Physics, University  of Colorado, Boulder, CO 80309}

\author{Douglas A. Bennett}
\affiliation{National Institute of Standards and Technology, Boulder, CO 80305}

\author{Shannon M. Duff}
\affiliation{National Institute of Standards and Technology, Boulder, CO 80305}

\author{Michael R. Vissers}
\affiliation{National Institute of Standards and Technology, Boulder, CO 80305}

\author{Joel N. Ullom}
\email{joel.ullom@nist.gov}
\affiliation{National Institute of Standards and Technology, Boulder, CO 80305}
\affiliation{Department of Physics, University  of Colorado, Boulder, CO 80309}

\maketitle

\onecolumngrid
\vspace{-1.5 cm}
\section{Device fabrication}
\begin{figure*}[ht!]
    \begin{subfigure}[b]{0.5\textwidth}
        \centering
        \caption{}
        \includegraphics[scale = 0.47]{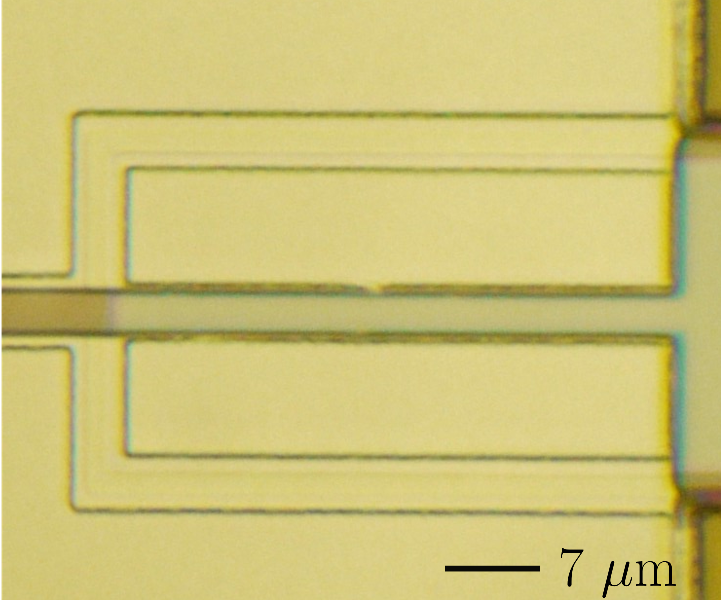}
    \end{subfigure}%
    ~ 
    \begin{subfigure}[b]{0.5\textwidth}
        \centering
        \caption{}
        \includegraphics[scale = 0.55]{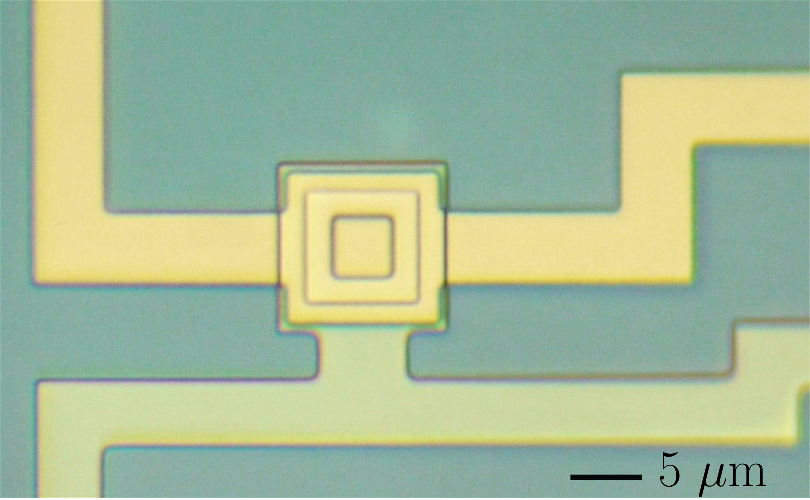}
    \end{subfigure}
    \caption{Optical microscope device pictures (a). SINIS AlMn \textbf{--}AlOx \textbf{--}Al junctions, (b). SIS   Nb \textbf{--} AlOx \textbf{--} Nb junction}
    \label{figS1}
\end{figure*}

The AlMn \textbf{--} AlOx \textbf{--} Al normal metal - insulator - superconductor (NIS) junctions were fabricated on a 3 inch diameter double-side polished (DSP) wafer where 120 nm thick thermal SiOx and 300 nm thick low-stress and low-pressure chemical vapor deposition (LPCVD) SiNx layers are  grown. A 30 nm AlMn base electrode is deposited by sputter deposition, patterned with photolithography, and etched with a heated aluminum etchant. The base electrode is passivated by a PECVD SiNx film. Vias are etched in the SiNx to form the area of the NIS junctions. The base electrode is cleaned with an ion mill, oxidized, and then capped with an Al superconducting layer of 300 nm. That layer is then oxidized and capped with a 500 nm AlMn normal metal layer which acts as a quasiparticle trap. The stack is then patterned and wet etched. This results in the SINIS structure shown in Fig. \ref{figS1}a where each NIS junction has an area of 32 $\times$ 7 $\mu \rm m^2$.

The Nb \textbf{--} AlOx \textbf{--} Nb superconductor-insulator-superconductor junctions were fabricated with a trilayer process. In vaccuum, 200 nm of niobium is deposited along with a thin layer of aluminum ($\sim 14$ nm), oxygen is introduced in the chamber to oxidize the aluminum, and thereafter, a 120 nm niobium layer is deposited on top. The counter and base electrodes of the trilayer are patterned using a combination of photolithography and reactive ion etching which defines the junction area. Then, an insulating SiO$_2$ film is deposited where vias are etched into the film to allow for electrical connectivity between additional Nb wiring layers and the junction. The junctions measured here had junction areas from 2.5 $\times$ 2.5 $\mu \rm m^2$ and 7.8 $\times$ 7.8 $\mu \rm m^2$ (shown in Fig. \ref{figS1}b).

\vspace{-0.45cm}

\section{Four-wire current voltage measurements}

\begin{figure}[ht!]
    \centering
    \includegraphics[scale = 0.4]{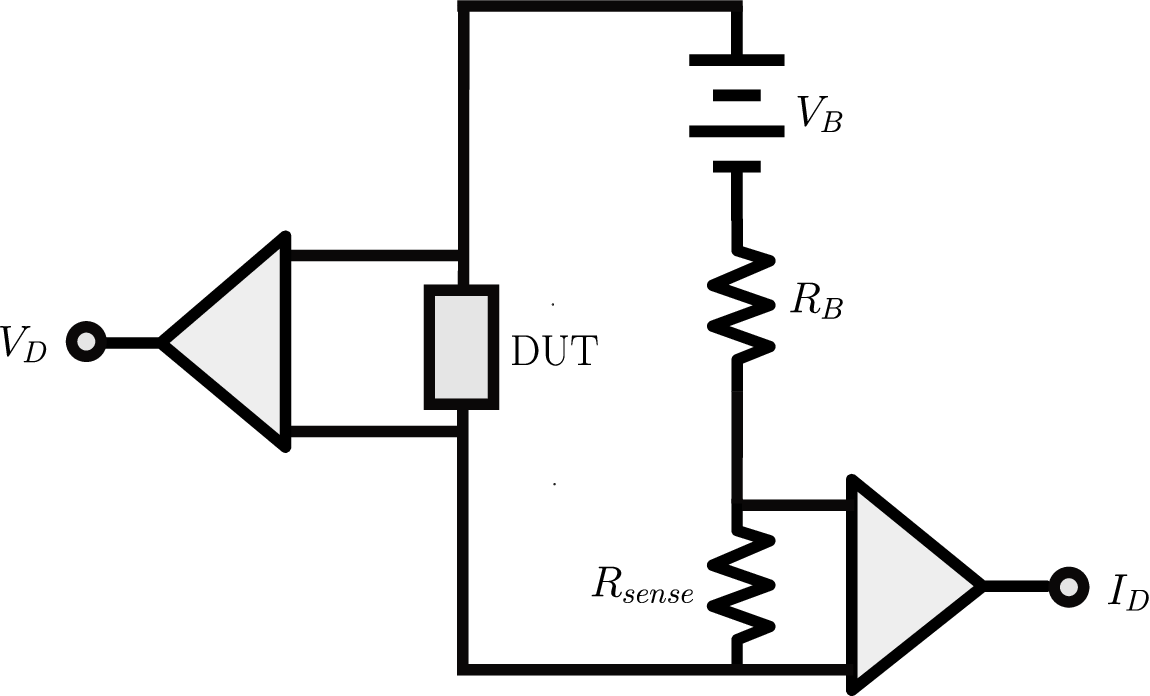}
    \caption{Electrical schematic for the four-wire current voltage measurement.}
    \label{figS2}
\end{figure}

Device measurements were performed in an adiabatic demagnetization refrigerator (ADR) with a base temperature of about 70 mK. A dual-polarity voltage is input into a bias resistor, $R_B$, which controls the current through the device and the current scale of the measurement. This current is converted into a voltage via a second resistor $R_{sense}$. The voltage across the device and $R_{sense}$ are then amplified by instrumentation amplifiers with a gain $G$ and recorded using digital multimeters. The measurement set-up is depicted in Fig. \ref{figS2}. When the device-under-test (DUT) is a high-resistance NIS junction, we take data by alternating between the positive and negative bias voltages so that we can subtract off potential drifts in the voltage and current measurement. Each of the measurement lines is filtered with 50 kHz low-pass filters at room temperature, 3 GHz and 80 MHz low-pass filters at 4 K, and 80 MHz low-pass filters at the base temperature. 

To deduce the actual current flowing through the device, $I_D$, from the amplified voltage across $R_{sense}$, $V_{m, I}$, we have
\begin{equation*}
    I_D = \frac{V_{m, I}}{R_{sense} G}.
\end{equation*}
And for the actual voltage across the device, $V_D$, from the amplified voltage, $V_{m,V}$,
\begin{equation*}
    V_D = \frac{V_{m,V}}{G}.
\end{equation*}

\begin{figure}[ht!]
    \centering
    \includegraphics[scale = 0.5]{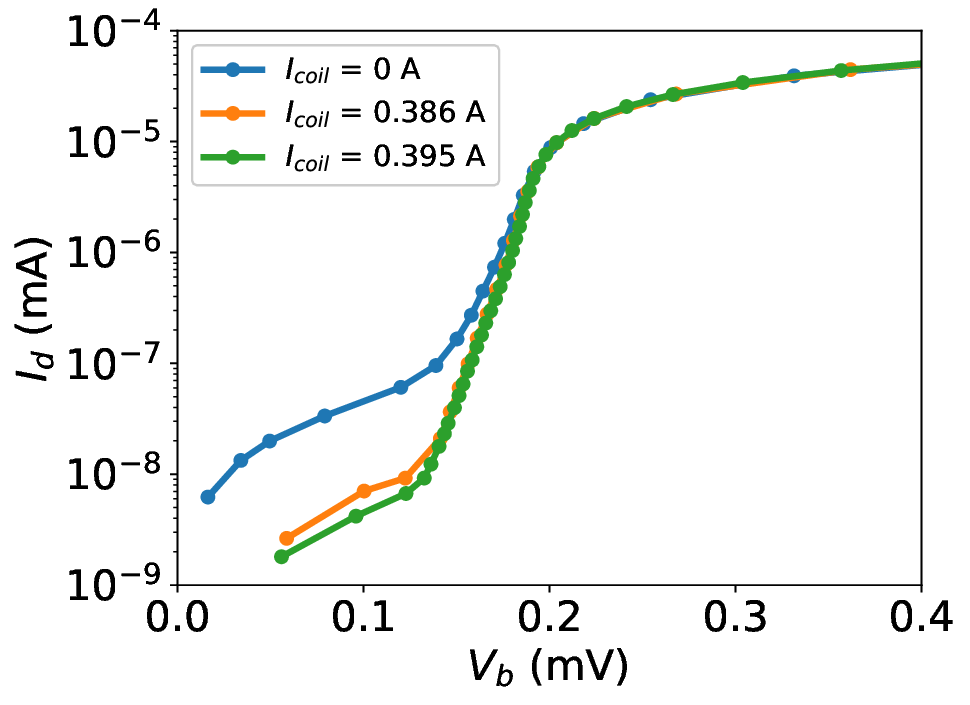}
    \caption{Effect of tuning the current through the external coil to minimize leakage current on the SINIS current-voltage curves.}
    \label{figS3}
\end{figure}

Furthermore, we use an external coil around the vacuum jacket of the ADR to optimize the magnetic field cancellation in our measurements. In particular, our NIS devices which had a larger area than the SIS junctions were susceptible to flux trapping which resulted in a higher leakage current. By tuning the current through the coil as the device passed through its critical temperature, we could achieve lower leakage currents. This effect is depicted in Fig. \ref{figS3}, where there is less leakage current as we tune the current through the coil and therefore the magnetic field.

\vspace{-0.45 cm}
\section{Dynes model fits to measured current-voltage curves}
In Figure 2 of the main text we show two representative fits to current-voltage data for aluminum and niobium junctions using the AG theory density of states and Lifshitz tail density of states. In this section, we show fits to the Dynes model which has been commonly used to describe sub-gap current leakage in tunnel junctions. The Dynes density of states is given as \cite{Dynes_1984}:
\begin{equation}
    \nu(E) = \rm{Re} \left ( \frac{E + i \Gamma}{\sqrt{ (E + i \Gamma)^{~2} - \Delta^2}} \right )
\end{equation}

\noindent where $\Gamma$ is the Dynes parameter. 
\begin{figure*}[ht!]
    \begin{subfigure}[b]{0.5\textwidth}
        \centering
        \caption{}
        \vspace{-0.2cm}
        \includegraphics[scale = 0.5]{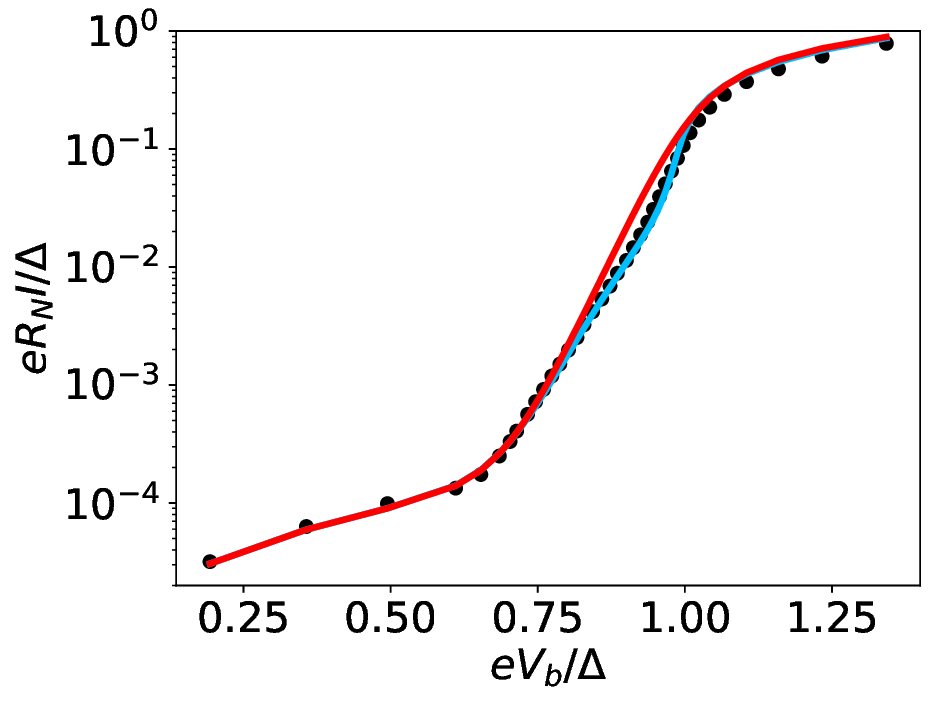}
    \end{subfigure}%
    ~ 
    \begin{subfigure}[b]{0.5\textwidth}
        \centering
        \caption{}
        \vspace{-0.2cm}
        \includegraphics[scale = 0.5]{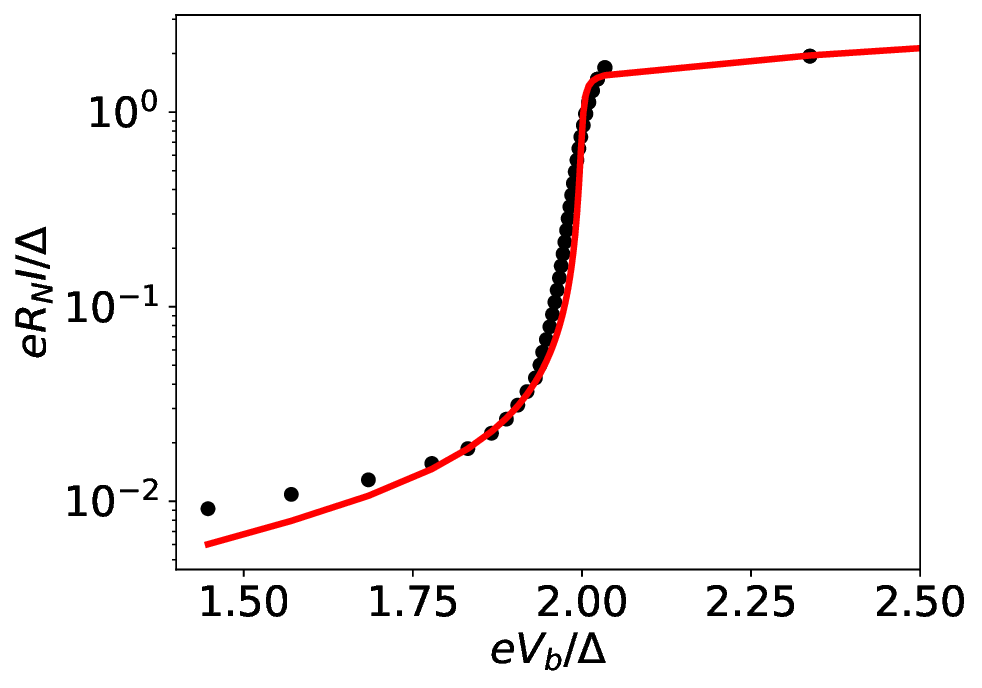}
    \end{subfigure}
    \caption{Measured data in black and representative fits with the Dynes model in red. (a). The AlMn \textbf{--} AlOx \textbf{--} Al NIS data is well fit with a Dynes Parameter $\Gamma / \Delta = 1.5 \times 10^{-4}$. Blue curve is the same Dynes model but including the self- cooling within the SINIS structure as described in \cite{ONeil_2012}. The coherence length in the dirty limit is $\xi \approx 800$ nm . (b) The  Nb \textbf{--} AlOx \textbf{--} Nb SIS data is fit with a Dynes parameter of $\Gamma / \Delta = 2.4 \times 10^{-3}$. The coherence length in the dirty limit is $\xi \approx 110 $ nm.  }
    \label{figS4}
\end{figure*}
The fits are done to the same data presented in the main text and are shown in Fig. \ref{figS4}. We observe that the Dynes model fits well for the AlMn \textbf{--} AlOx \textbf{--} Al NIS junction and less so for the  Nb \textbf{--} AlOx \textbf{--} Nb SIS junction. Nonetheless, the Dynes parameters we see in our data are typical of what has been observed in the literature, suggesting that photon-assisted tunneling may be responsible for the sub-gap leakage \cite{Pekola_2010}.

\begin{table}[ht!]
    \caption{Summary of key parameters ($\varepsilon_g / \Delta$, $\varepsilon_t / \Delta$) extracted from fitting DOS of the measured devices, and calculated localization radius, $r_c$. The ranges of $n_{qp,c}$ are calculated using $\varepsilon_{t,LO}$ (yields lower value) and $\varepsilon_{t,MS}$ (yields upper value).}

    \centering
    \begin{tabularx}{1.02 \textwidth}{| >{\centering\arraybackslash}X 
  | >{\centering\arraybackslash}X 
  | >{\centering\arraybackslash}X
  | >{\centering\arraybackslash}X
  | >{\centering\arraybackslash}X
  | >{\centering\arraybackslash}X
  | >{\centering\arraybackslash}X
  | >{\centering\arraybackslash}X |}
        \hline
        \textbf{Device} & $\Gamma / \Delta$ & $\varepsilon_g / \Delta$ & $\varepsilon_{t,MS} / \Delta$ & $r_{c,MS}$ (nm) &$\varepsilon_{t,LO} / \Delta$ & $r_{c,LO}$ (nm) & $n_{qp,c} ~ (\mu\rm m^{-3})$ \\
        \hline  \hline
        Al NIS 1 & $1.5 \times 10^{-4}$ & $2.75 \times 10^{-2}$ &  $2.39 \times 10^{-2}$ & 409 & $2.27 \times 10^{-2}$ & 553 & $(0.03 - 0.07)$ \\
        \hline
       Al NIS 2 & $2.2 \times 10^{-4}$  & $2.97 \times 10^{-2}$  & $2.41 \times 10^{-2}$  & 410 & $2.31 \times 10^{-2}$ & 553 & $(0.03 - 0.07)$ \\
        \hline
       Al NIS 3 & $1.6 \times 10^{-4}$ & $3.37 \times 10^{-2}$ &  $2.52 \times 10^{-2}$ & 401 & $1.5 \times 10^{-2}$ & 614 & $(0.03 - 0.08)$ \\
        \hline \hline
        Nb SIS 1 & $2.4 \times 10^{-3}$ & $2.02 \times 10^{-2}$ & $1.37 \times 10^{-2}$ & 79 &  $7 \times 10^{-3}$ & 121 & $(3 - 10)$ \\
         \hline
         Nb SIS 2 & $3.8 \times 10^{-3}$  & $3.28 \times 10^{-2}$ & $2.48 \times 10^{-2}$ & 70 & $1.28 \times 10^{-2}$ & 105 & $(4 - 14.5)$  \\
         \hline
         Nb SIS 3 & $7 \times 10^{-3}$  & $6.54 \times 10^{-2}$ & $4.42 \times 10^{-2}$ & 61 & $2.87 \times 10^{-2}$& 89 & $(7 - 22)$\\
         \hline
    \end{tabularx}
    \label{tabS1}
\end{table}

\vspace{-0.45cm}
\section{Summary of other device measurements}
In Table \ref{tabS1}, we show the fit results for all of the aluminum NIS and niobium SIS junctions that we measured, along with the extracted parameters. We can see that our measurements yield similar values for our parameters of interest $\varepsilon_g / \Delta$ and $\varepsilon_t / \Delta$ for each type of junction.

\bibliographystyle{unsrtnat}
\bibliography{supp_references.bib}